\documentclass{article}

\usepackage{amssymb}
\usepackage{amsmath}
\usepackage{amsfonts}
\usepackage{amssymb}
\usepackage{graphicx}
\usepackage{color}%
\usepackage[utf8]{inputenc}
\usepackage{color}%
\usepackage[normalem]{ulem} 
  \usepackage{xr}
\usepackage[authoryear]{natbib}
\usepackage{hyperref}

\title{ Each student with her/his own data: understanding sampling distributions}
\author{Mariela Sued$^{(1)}$, Marina Valdora$^{(1)}$}

\begin{document}
\maketitle
{\small (1)  Universidad de Buenos Aires and Conicet.}

\section*{Abstract}

{Sampling distribution,   a  foundational  concept in statistics, is  difficult to understand, since we usually have only one realization  of the estimator of interest. In this work, we present an innovative method for helping university students understand 	
	 the variability of an estimator. In our approach,  each student uses a  different data set, getting diverse  estimations.  Then, sharing the results, we can empirically study the sampling distribution. 

After some \textit{handmade} experiences, we have built a web page to deliver a personalized data set for each student. 
	Through this web page, we can also reformulate the role of the student in the classroom, inviting him/her to became an active player, submitting the solution to different problems and checking whether they are correct. 
	In this work we  present a recent experience in such direction.
}

\section{Introduction}
A typical problem for users of statistics is to assess the variability of an estimator. As an example, a couple of years ago, someone came to us with the following problem: 

\lq\lq
We reported a median of 
$3.21$ but the referee asked for the error. What should we do? I brought you this. \rq\rq And they gave us the information included in  Figure \ref{fig:hist_datos}: a histogram, the median (3.21), the mean (4) the standard deviation of the data (3.40), and en the sample size ($n=120$).
\lq\lq For the mean, we know it. It is $3.40/\sqrt{120}$ \rq\rq.
As far as we can see, quantifying the variability of an estimator other than the mean, can usually be counted among these problems.
What is going on? Besides knowing the formula for computing the standard error of the mean, do students really understand what it represents?
The notion of variability associated to data is natural: we have many different values, we can make histograms, boxplots, etc. But what matters to the referee is the variability of the estimator, {the sampling distribution},  for which we only have one {\bf realization}. 

{Sampling distribution is one of the most important  concept in statistics, fundamental for inference procedures, like confidence intervals and p-values.  Teaching sampling distributions   is one of biggest  challenges to be faced in the classroom. Many authors have  discussed and written  about this issue in recent years. For instance, \cite{hancock2020simulation} have recently  described an empirical intervention which shows that 		hands-on activity before computer simulation are useful to gain understanding on  sampling distribution. In their work the authors present a complete review on the existing literature on the subject. See  also \cite{garfield2007students} and \cite{sotos2007students}.

Before going on,  we would like to differentiate between two kinds of students: those who have experience working in a laboratory (students of chemistry, biology and physics) and those who have not (students of mathematics and computer science, among others).
We 
shall call the former {\it{lab students}}.

After years of teaching, we have realized that lab students are more familiar with the notion of sampling distribution; they 
know that each of the {students} in the lab obtains a different result based on their own data set. 
Even though each lab student obtains only one estimation, they know that the student next to them gets a different estimation and so does each member of the class. If they gathered all the estimates, they could do descriptive statistics in order to understand the so-called sampling distribution, that is to say, the distribution of the estimator they are using. Lab students also know that, when reporting the value of an estimate, it is important to include an \lq\lq error \rq\rq. Typically, {they} write {\it{estimate (error) }}, where error is an estimation of  the standard deviation of the sampling distribution of the estimator. For instance, considering the values given above, when they want to inform the mean and its error,  they would express it as $0.31 (3.40/\sqrt{120})$.


The last statement is a consequence of the fact that mathematics gives us a formula for the standard deviation of the mean: $\sigma/\sqrt{n}$, where $\sigma$ is the standard deviation of the data generating distribution and $n$ is the sample size. Thanks to this formula,  it is easy to report errors when the estimator is the mean: one simply replaces $\sigma$ by an estimate such as \verb|sd(data)|, and thus \verb|error=sd(data)/sqrt(n)|. 
This friendly formula makes sense even without a deep understanding of its origin. It is related to the variability of the data and decreases when $n$ increases.

Even though the notion of sampling distribution is clear for lab students, it may not be so for those who have no experience in the lab, even those who have already worked with real data because they only observe one data set and therefore they only have one estimate. Our classes do not help much with this issue; 
usually our exercises are to be solved based on a data set that is the same for all students.
For example, 

{\textit{ A set of 20 lamps from a certain manufacturer has been tested and the following lifetimes have been obtained (in days).}}

\begin{tabular}{rrrrrrrrrr}
	39.08 & 45.27 & 26.27 & 14.77 & 65.84 & 49.64 & 0.80 & 66.58 & 69.60 & 32.42 \\ 
	228.36 & 64.79 & 9.38 & 3.86 & 37.18 & 104.75 & 3.64 & 104.19 & 8.17 & 8.36 \\ 
\end{tabular}

{\textit{
Based on these observations, estimate the mean and the median 
of the lifetime of a lamp  produced by this manufacturer and the error associated to each estimator.}
}

A good student would do the following:
\begin{verbatim}
> mean_estimate <- mean(data)
> mean_estimate
[1] 49.1475
> mean_error <- sd(data)/sqrt(length(data))
> mean_error
[1] 11.82242
> median_estimate <- median(data)
> median_estimate
[1] 38.13
\end{verbatim}
and he would come to ask about the median error. 
Before answering this question, we  want to make sure  they understand the concept of estimation error and, more generally, sampling distribution.
In the following sections we will present our proposal to attain this goal.

\section{Our proposal}

In order to help the students incorporate the concept of {\itshape sampling distribution}, we invite teachers to replace the previous type of exercise by a new one in which each student will work based on her own data. The exercise begins with teachers delivering a data set for each student (or group of students), with the possibility of changing the sample size. Below, we discuss a form of doing this. Then, the students compute the estimates of interest based on samples of different sizes. Finally, the teacher collects all the estimates computed by the class, does descriptive statistics for each sample size and comments on the results. This kind of activity can be presented at different instances of a statistics course, allowing the teacher to emphasize concepts of interest at each stage. 

In our first implementations of this proposal, we began by printing different data sets in pieces of paper and handing them out in class.
After a successful test trial, we  decided to invest time in creating shiny apps (\citep{shiny}).
Shiny apps are a versatile tool that, so far, have allowed us to:  
\begin{itemize}
	\item deliver a different data set to each student, using a personal id number as a seed. 
	\item emulate the process of increasing the sample size by adding more data to the original set.
	\item give students the possibility to check their answers. 
\end{itemize}

\section{An example with shiny}
The exercise above should be replaced by 

{\itshape You want to study the distribution of the lifetime (in days) of the lamps produced by a certain company. With this aim you select $n$ lamps from its production, try them and register their lifetime. {To see the results of your experiment for different values of $n$ }  click
\href{https://marin.shinyapps.io/Lifetimes2/}{here}.
\begin{enumerate}
	\item[a)] For $n=5$, $n=30$ and $n=100$, compute the mean and median lifetime of your data set and check your results following the instructions on the app.
	\item[b)] Once you  verified the correct answers, for each value of $n$ fill the answers in the following
	\href{https://docs.google.com/forms/d/e/1FAIpQLSeiRJ3LiuA6h7HO7b2VaJNIWefwGZ3SHRntB6CETO2htdoOnQ/viewform?usp=sf_link}{ form}.
\end{enumerate} 
}

In this way a file will be generated, containing one mean and one median lifetimes for each sample size (columns) and for each student (rows).

Figure \ref{fig:captura1} is a screenshot of the webpage the previous link directs you to.

\subsection{Working with the mean}\label{sec:mean}

In class, we visualize the means reported by the students for each sample size and invite them to identify their results.
Then we draw probability histograms with the means for each sample size with the same scale in the x axis. {Histograms for each required sample size ($n=5,30,100$) resulting from a class with $80$ students are given in Figure \ref{fig:hist_means}.}

The students are guided to discover that the means are more concentrated as $n$ increases. 

Then we discuss the relation between the standard deviation of the means for each sample size and the error reported by each student. 

It should be clear that for large $n$  the error reported by the students are very similar and they are also similar to the standard deviation of the reported means for the same $n$.

This activity should show that the standard error is nothing but an estimation of the standard deviation of the sampling distribution.
\subsection{Working with the median}

We repeat the work done for the mean, except that now the students are not able to report an error. {Histograms of the medians for each required sample size ($n=5,30,100$) resulting from a class with $80$ students are given in Figure \ref{fig:hist_medians}.}

At this point we resume the discussion from Section \ref{sec:mean}, where we concluded that the error is nothing but an estimation of the standard deviation of the sampling distribution.  This is the definition of error and it applies to any estimator; in particular, the median.
From Wikipedia:

{\itshape
\textbf{Definition} The standard error of an estimation is the standard deviation (or an estimation) of the distribution of the estimator.}

Based on this definition and the 80 medians obtained by the class, we can compute the standard error of the median for each sample size by the empirical standard deviation of the set of medians, namely, if \verb|median_5| stores the 80 medians obtained by the students in the class with sample size $n=5$, \verb|sd(median_5)| can be considered as the required standard error.

In real life, we only have one data set and therefore only one estimation. What we have done here is unfeasible because we cannot build a histogram with just one datum. However, after doing the type of exercise we have proposed, students will hopefully understand the notion of sampling distribution and standard error.

\section{Final remarks}
 The next question is, of course, how to obtain the sampling distribution and the standard error when we only have one data set. Mathematical statistics proposes different methods to approach this problem, most of them based on the central limit theorem. On the other hand, an alternative, computer-based proposal is non-parametric wild bootstrap, which is a way to emulate what we have done in our proposed exercise but with only one data set. We hope that this kind of classroom work sets the foundations for understanding resampling methods.

{Before concluding,  we want to emphasise that the app used to illustrate  our proposal is far from perfect. Since we are not experts in shiny, we understand that this is barely a prototype, than can be largely improved. For this reason, we decided to share our experience with the community and invite everyone to create new challenges, making use of the enormous power of Shiny or other applications that can help to preach our message: individual data sets and active role when interacting with web  applications.} 

{This manuscript was finished while the coronavirus exploded around the word. 
In-person classes have been recently cancelled and teachers have been invited 
to implement virtual teaching. In this context, we trust  that our proposal may help 
to produce interesting activities.}

\bibliographystyle{plain}
\bibliography{shiny}

\begin{figure}\centering
	\includegraphics[scale=0.6]
	{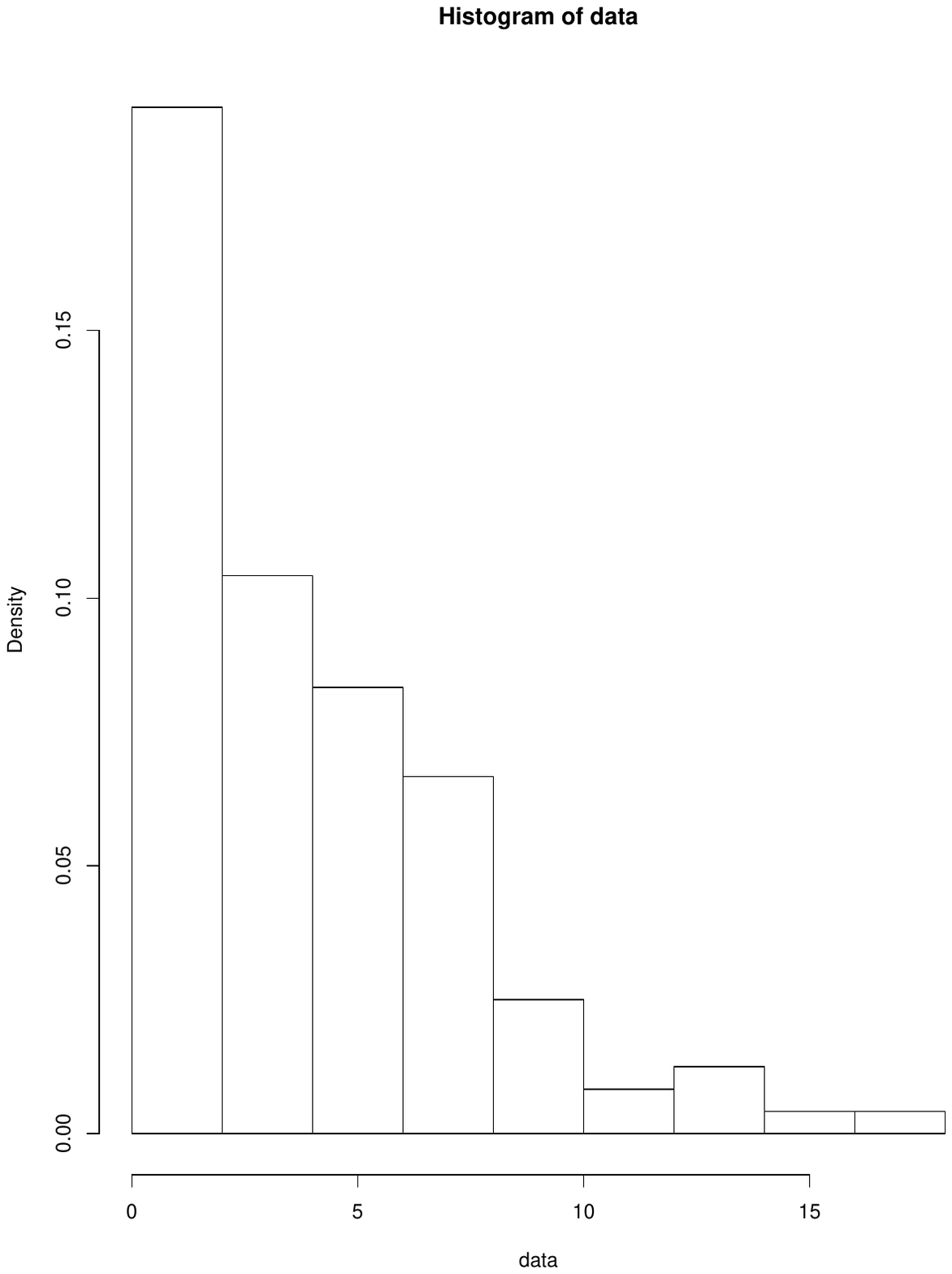} 
	\caption{Histogram, median=$3.21$, mean$=4$, sd$=3.40$} and $n=120$. \label{fig:hist_datos}
\end{figure}

\begin{figure}[h]
	\centering
	\includegraphics[scale=0.35]{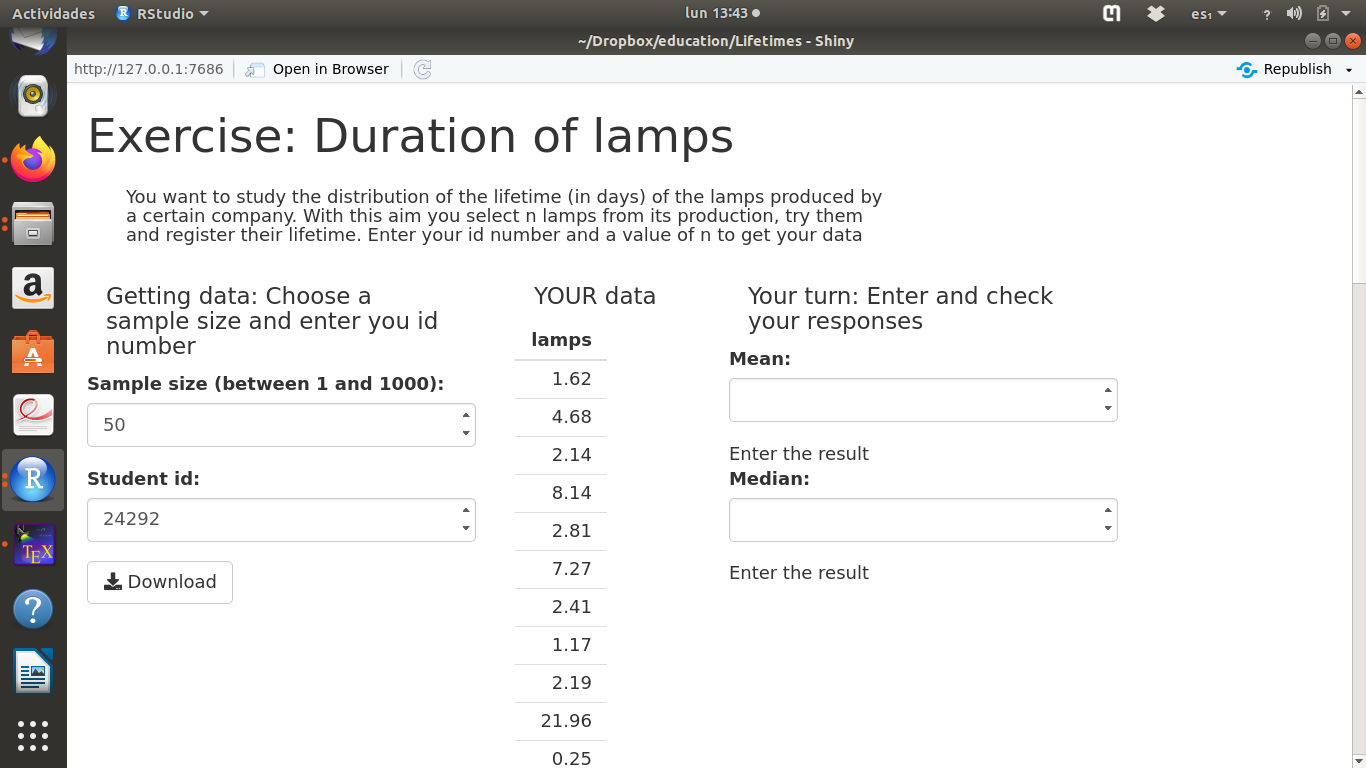}
	\caption{}
	\label{fig:captura1}
\end{figure}

\begin{figure}[h]
	\centering
	\includegraphics[scale=0.8]
	{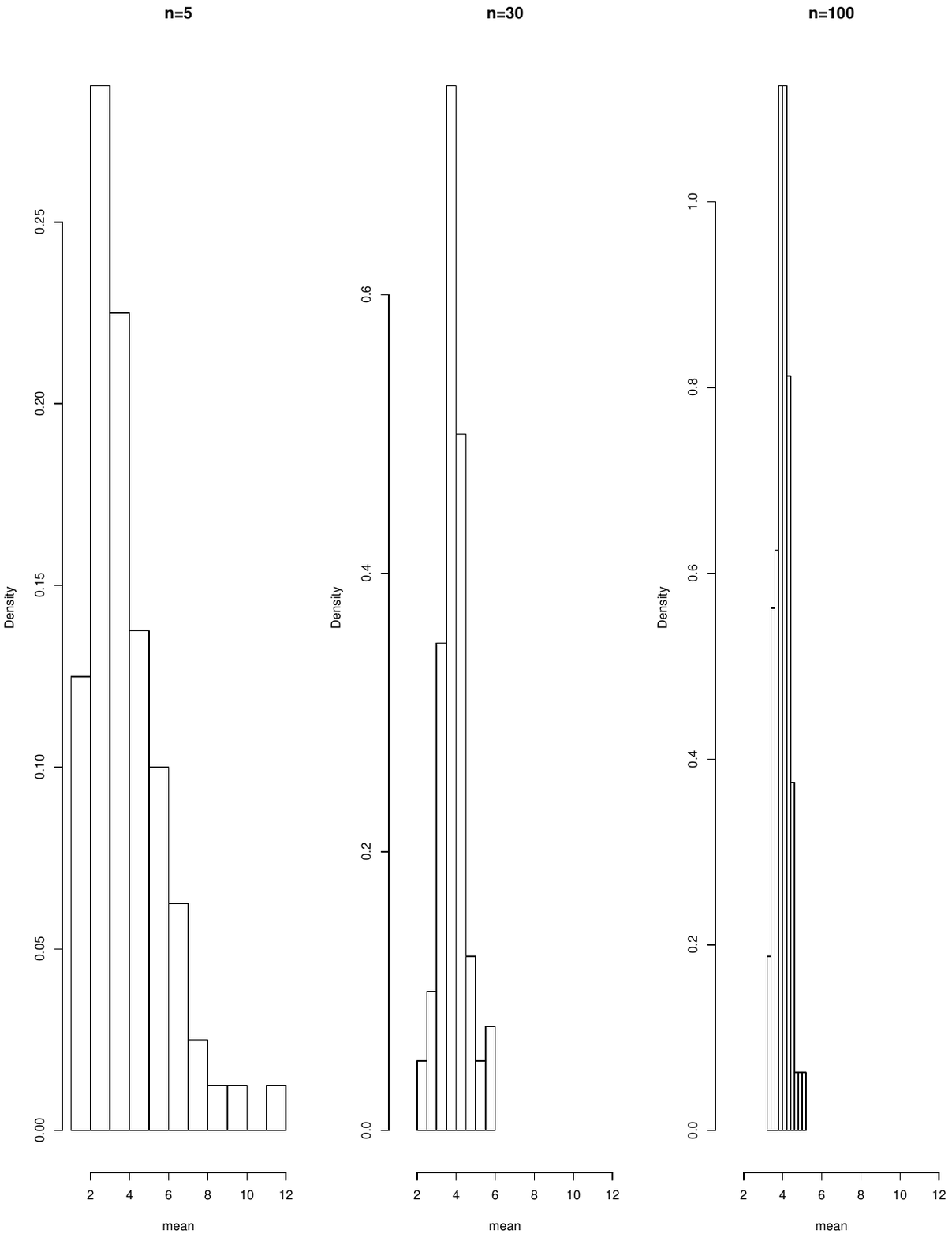}
	\caption{Histograms of means for different values of $n$}
	\label{fig:hist_means}
\end{figure}

\begin{figure}[h]
	\centering
	\includegraphics[scale=0.8]
	{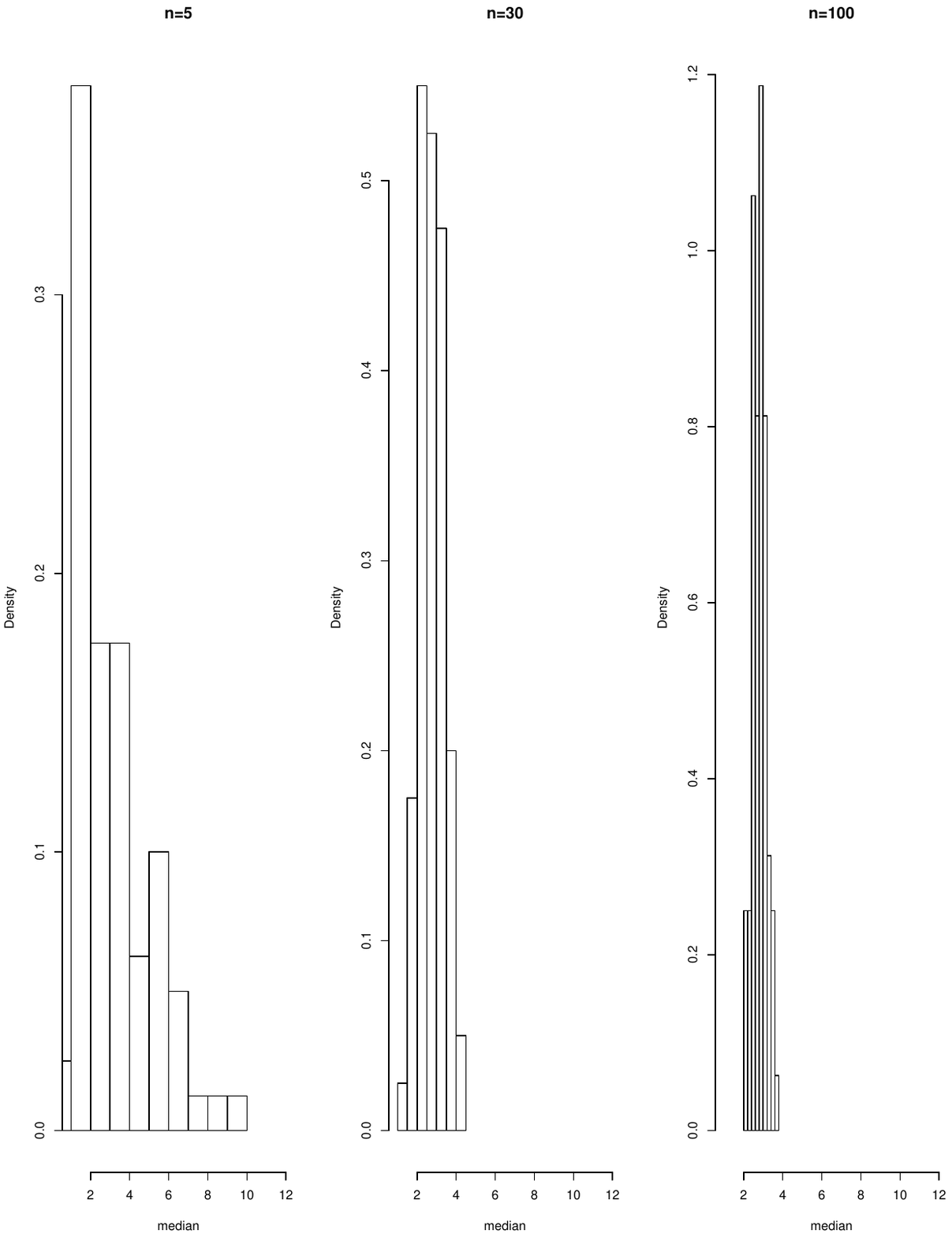}
	\caption{Histograms of medians for different values of $n$}
	\label{fig:hist_medians}
\end{figure}

\end{document}